# visionFinGAR: Transmission of Softness and Shape Motion by Vision Based Tactile Sensor and Combination of Mechanical and Electrical Stimulation

Hikaru Kukita[1], Hiroyuki Kajimoto[2], and Vibol Yem[1]

[1] The University of Tsukuba, Tsukuba, Japan
[2] The University of Electro-Communications, Tokyo, Japan
(Email: h_kukita@vrlab.esys.tsukuba.ac.jp, kajimoto@uec.ac.jp, yem@iit.tsukuba.ac.jp )

**Abstract ---** This paper describes a system for transmitting softness and the motion of shape or contact area sensation using a vision based tactile sensor and a tactile display in which mechanical and electrical stimulation are combined. A unit of tactile sensor consists of a camera and markers, enable to detect a light touch, a pressure or a shape. On the other hand, a unit of tactile display consists of an electrode array and a mechanical arm to provide softness / pressure and shape perception. The display can provide four mode stimulation: anodic, cathodic, mechanical vibration and skin deformation; thus, it can reproduce a large range of tactile sensations. This study mainly aims to transmit a wide range of softness and shape motion perception with a vision based tactile sensor.

## 1 INTRODUCTION

Technological advancements in communication have enabled remote sharing of voice and vision with high quality. However, transmitting the sensation of touch remains a challenge. While tactile transmission is crucial for tele-operation, tele-training, and telexistence [1], current methods struggle to provide realistic tactile feedback due to limitations in tactile sensor capabilities and tactile display technology.

Existing tactile sensors, like accelerometers, microphones, and pressure sensors, have limited spatial or temporal resolution. On the other hand, tactile displays require high spatial and temporal resolution to reproduce a wide range of tactile sensations. Despite the potential to stimulate all relevant tactile receptors with sufficient resolution, developing a versatile micro-machine for tactile displays remains a challenge due to the skin's inherent mass and damping properties. In our previous study, we developed both a tactile sensor and a tactile display with high spatiotemporal resolution. The sensor, integrated into a glove, measures temperature and pressure distribution across three fingers. The display, also embedded in a glove, utilizes electrotactile, vibrotactile, and thermal actuators to reproduce the measured tactile sensations [2]. However, the sensor was not suitable to detect a light touch, and the display cannot generate skin deformation or very-low frequency vibration; thus, it was difficult to transmit softness and shape motion (the motion of contact area) perception.

In our present study, we use a vision-based sensor [3] to detect light touch as well as pressure and shape. We also use the combination of mechanical and electrical stimulation with four modes: anodic, cathodic, mechanical vibration and skin deformation, to provide a wide range of tactile sensation. Compared to our previous device presented in [4], the current device has a higher dimensional electrode array; therefore, it can provide shape motion perception clearer.

## 2 SYSTEM

### 2.1 Vision-based tactile sensor

FingerVision (FingerVision Inc.) is used to measure touch perception (Fig. 1-a). FingerVision, a novel tactile sensing system, combines a transparent, elastic skin marked with dots with a simple and robust camera. It can detect force distributions, slip, and object properties like distance, location, pose, size, shape, and texture. FingerVision enables to perform complex manipulations, including grasping delicate objects like origami, vegetables, fruits, and raw eggs. Therefore, we consider that it is also suitable for softness and shape motion detection.

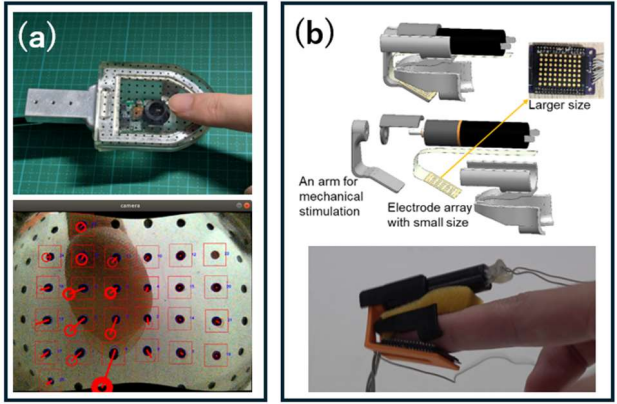

Fig.1 (a) Vision-based sensor for light touch, pressure and shape motion detection. (b) Design of tactile display to present mechanical and electrical stimulation [4].

### 2.2 Tactile display with four mode stimulation

A tactile feedback device (FinGAR) was designed for virtual and augmented reality applications, aims to provide realistic touch sensations to multiple fingers (Fig. 1-b). To achieve this, FinGAR is designed as a lightweight, wearable device that can be attached to the thumb, index finger, and middle finger, without hindering natural finger movement. Each FinGAR unit consists of five components: a 3D-printed ABS finger glove, a DC motor, a DC motor stopper, an arm, and an electrode film. The finger glove securely grips the finger, while the other components work together to deliver four-mode tactile stimulation: anodic and cathodic by electrode array, and vibration and skin deformation by a mechanical arm. This device was published in [4], and this present device increases the size of electrode array.

### 3 DEMONSTRATION

Fig. 2 shows the overview of system demonstration using our visionFinGAR. In the demonstration, we use various shapes and softness of the real objects touching on the sensor and users can experience those shapes and softness sensation via the stimulation of our tactile display.

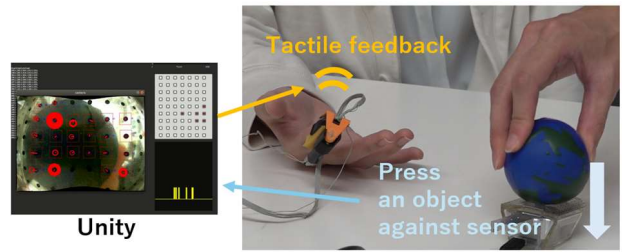

Fig.2 System demonstration using visionFinGAR

### 4 CONCLUSION

This study advances tactile communication by enabling the transmission of a broader range of sensations, including light touch, pressure, shape motion, and softness. Building upon our prior work, we integrated a vision-based sensor for detecting light touch and pressure and developed a multi-modal stimulation system with four modes (anodic, cathodic, mechanical vibration, and skin deformation). This enhancement allows for more realistic tactile experiences, particularly in transmitting softness perception. The increased size of the electrode array also improves shape motion perception. These advancements pave the way for more immersive telepresence and remote manipulation technologies.

ACKNOWLEDGEMENT

Research supported by JST A-STEP Grant Number JPMJTR23RC, Japan.